\documentclass[10pt,conference]{IEEEtran}
\usepackage{maxim}
\usepackage{graphicx}
\usepackage{subfigure}
\usepackage{verbatim}

\def\cA{{\cal A}}

\def\cC{{\cal C}}

\def\cN{{\cal N}}

\def\cS{{\cal S}}

\def\cX{{\cal X}}

\def\cZ{{\cal Z}}

\def\sV{{\sf V}}

\def\hcX{\widehat{\cal X}}
\def\hX{\widehat{X}}
\def\hx{\widehat{x}}

\def\dvar{d}

\def\bd#1{\boldsymbol{#1}}
\def\wh#1{\widehat{#1}}
\def\td#1{\widetilde{#1}}

\begin{document}

\title{Joint Universal Lossy Coding and Identification\\
of Stationary Mixing Sources}

\author{\authorblockN{Maxim Raginsky}
\authorblockA{Beckman Institute and the University of Illinois\\
405 N Mathews Ave, Urbana, IL 61801, USA \\
Email: maxim@uiuc.edu}}
\maketitle

\begin{abstract}
The problem of joint universal source coding and modeling, treated in the context of lossless codes by Rissanen, was recently generalized to fixed-rate lossy coding of finitely parametrized continuous-alphabet i.i.d. sources. We extend these results to variable-rate lossy block coding  of stationary ergodic sources and show that, for bounded metric distortion measures, any finitely parametrized family of stationary sources satisfying suitable mixing, smoothness and Vapnik--Chervonenkis learnability conditions admits universal schemes for joint lossy source coding and identification. We also give several explicit examples of parametric sources satisfying the regularity conditions.
\end{abstract}

\section{Introduction}
\label{sec:intro}

A universal source coding scheme is one that performs asymptotically optimally for all sources within a given class. Intuition suggests that a good universal coder should acquire a probabilistic model of the source from a sufficiently long data sequence and operate based on this model. For lossless codes, this intuition has been made rigorous by Rissanen \cite{Ris84}: the data are encoded via a two-part code which comprises (1) a suitably quantized maximum-likelihood estimate of the source parameters, and (2) an encoding of the data with the code optimized for the acquired model. The redundancy of this scheme converges to zero as $k\log n/n$, where $n$ is the block length and $k$ is the dimension of the parameter space.

Recently we have extended Rissanen's ideas to {\em lossy} block coding  of finitely parametrized continuous-alphabet i.i.d. sources with bounded parameter spaces \cite{Rag05,Rag06}. We have shown that, under appropriate regularity conditions, there exist joint universal schemes for lossy coding and source identification whose distortion redundancy and source estimation fidelity both converge to zero as $O\big(\sqrt{\log n/n}\big)$ as the block length $n$ tends to infinity. The code operates by coding each block with the code matched to the parameters estimated from the preceding block. Moreover, the constant hidden in the $O(\cdot)$ notation increases with the ``richness" of the model class, as measured by the Vapnik--Chervonenkis (VC) dimension \cite{DevLug01,Vid03} of a certain class of decision regions in the source alphabet.

The main limitation of the results of \cite{Rag05,Rag06} is the i.i.d. assumption, which excludes such practically relevant model classes as autoregressive sources or Markov and hidden Markov processes. Furthermore, the assumption of a bounded parameter space may not be always justified. In this paper we relax both of these assumptions. Because the parameter space is not bounded, we have to use variable-rate codes with countably infinite codebooks, whose performance is naturally quantified by Lagrangians \cite{ChoEffGra96,Lin01}. We show that, under certain regularity conditions, there are universal schemes for joint lossy source coding and modeling such that, as the block length $n$ tends to infinity, both the Lagrangian redundancy relative to the best variable-rate code at each block length and the source estimation fidelity at the decoder converge to zero as $O( \sqrt{ V_n \log n / n})$, where $V_n$ is the VC dimension of a certain class of decision regions induced by the collection of all $n$-dimensional marginals of the source process distributions.

The key novel feature of our scheme is that, unlike most existing schemes for universal lossy coding, which rely
on implicit identification of the active source, it learns an explicit probabilistic model. Moreover, our results clearly show that the ``price of universality" of a modeling-based compression scheme grows with the combinatorial richness of the underlying model class, as captured by the VC dimension sequence $\{V_n\}$. The richer the model class, the harder it is to learn, which in turn affects the compression performance because we use the source parameters learned from past data in deciding how to encode the current block. These insights may prove useful in such settings as digital forensics or adaptive control under communication constraints, where trade-offs between the quality of parameter estimation and compression performance are of central importance.\begin{comment}Furthermore, comparing the rate at which the Lagrangian redundancy decays to zero under our scheme with the $O(\log n/n)$ result of Chou, Effros and Gray \cite{ChoEffGra96}, whose universal scheme is not aimed at identification, we immediately see that, in ensuring to satisfy the twin objectives of compression and modeling, we inevitably sacrifice some compression performance.\end{comment}

\section{Preliminaries}
\label{sec:prelims}

Let $\bd{X} = \{X_i\}_{i \in \Z}$ be a stationary, ergodic source with alphabet $\cX$. All alphabets are assumed to be Polish spaces equipped with their Borel $\sigma$-fields. We adopt the usual setting of universal source coding: the process distribution of $\bd{X}$ is not known exactly, apart from being a member of some indexed class $\{P_\theta : \theta \in \Lambda\}$. We assume that the parameter space $\Lambda$ is an open subset of $\R^k$ with nonempty interior. We also assume that there exists a $\sigma$-finite measure $\mu$ on $\cX$, such that for every $\theta \in \Lambda$ the $n$-dimensional marginals $P^n_\theta$ of $P_\theta$ are absolutely continuous with respect to (w.r.t.) the product measure $\mu^n$, for all $n$, denoting the corresponding densities $dP^n_\theta/d\mu^n$ by $p^n_\theta$.

We wish to code $\bd{X}$ into a reproduction process $\wh{\bd{X}} = \{\hX_i\}_{i \in \Z}$ with alphabet $\hcX$ by means of a finite-memory variable-rate lossy block code (vector quantizer). Such a code with block length $n$ and memory length $m$ [an $(n,m)$-block code, for short] is a pair $C^{n,m} = (f,\varphi)$, where $\map{f}{\cX^n \times \cX^m}{\cS}$ is the encoder, $\map{\varphi}{\cS}{\hcX^n}$ is the decoder, and $\cS \subseteq \{0,1\}^*$ is a finite or countable collection of binary strings satisfying the prefix condition. The mapping of $\bd{X}$ into $\wh{\bd{X}}$ is defined by $\wh{X}^{n(k+1)}_{nk+1} = \varphi(f(X^{n(k+1)}_{nk+1},X^{nk}_{nk-m+1}))$, $k \in \Z$, where $X^j_i \deq (X_i,X_{i+1},\ldots,X_j)$, $i < j$. Thus, the encoding is done in blocks of length $n$, but the encoder is also allowed to view the $m$ source symbols immediately preceding the current $n$-block. Abusing notation, we shall denote by $C^{n,m}$ both the composition $\varphi \circ f$ and the pair $(f,\varphi)$; when $m=0$, we shall use a more compact notation $C^n$ and say ``$n$-block code."

Let $\map{\rho}{\cX \times \hcX}{\R^+}$ be a measurable single-letter distortion function; $\rho_n(x^n,\hx^n) = n^{-1}\sum^n_{i=1}\rho(x_i,\hx_i)$ is the per-letter distortion due to reproducing $x^n \in \cX^n$ by $\hx^n \in \hcX^n$. We assume that $\rho$ is a metric on $\cX \cup \hcX$, bounded from above by some $\rho_{\max} < \infty$. Suppose $\bd{X} \sim P_\theta$. Associated with the code $C^{n,m}$ are its expected distortion $D_\theta(C^{n,m}) \deq \E_\theta \{\rho_n(X^n_1,\hX^n_1) \}$ and its expected rate $R_\theta(C^{n,m}) \deq \E_\theta \{\ell_n(f(X^n_1,X^0_{-m+1}))\}$, where, for a binary string $s$, $\ell_n(s)$ is its length in bits, normalized by $n$. When working with variable-rate quantizers, it is convenient \cite{ChoEffGra96,Lin01} to absorb the distortion and the rate into a single performance measure, the {\em Lagrangian} $L_\theta(C^{n,m},\lambda) \deq D_\theta(C^{n,m}) + \lambda R_\theta(C^{n,m})$, where $\lambda > 0$ is the {\em Lagrange multiplier} which controls
the distortion-rate trade-off. The optimal Lagrangian performance achievable on $P_\theta$ by any zero-memory variable-rate quantizer with block length $n$ is given by the {\em $n$th-order operational distortion-rate Lagrangian} $\wh{L}^n_\theta(\lambda) \deq \inf_{C^n} L_\theta(C^n,\lambda)$ \cite{ChoEffGra96}. Allowing the codes to have nonzero memory does not improve optimal performance, because we can use memoryless nearest-neighbor encoders to convert any $(n,m)$-block code into an $n$-block code without increasing the Lagrangian. Thus, $\wh{L}^n_\theta(\lambda) = \inf_m \inf_{C^{n,m}} L_\theta(C^{n,m},\lambda)$, where the infimum is over all memory lengths $m$ and all $(n,m)$-block codes $C^{n,m}$, for a fixed block length $n$. Because each $P_\theta$ is ergodic, $\wh{L}^n_\theta(\lambda)$ converges, as $n \to \infty$, to the {\em distortion-rate Lagrangian} $L_\theta(\lambda) \deq \min_R \Big(D_\theta(R) + \lambda R \Big)$, where $D_\theta(R)$ is the Shannon distortion-rate function of $P_\theta$ \cite{ChoEffGra96}. 

\section{The results}
\label{sec:result}

In this section we state our result on universal schemes for joint lossy compression and identification of stationary sources satisfying certain regularity conditions. We wish to design a sequence of variable-rate vector quantizers, such that the decoder can reliably reconstruct the source sequence $\bd{X}$ and reliably identify the active source in an asymptotically optimal manner for all $\theta \in \Lambda$. The identification performance will be judged in terms of the variational distance, which for any two probability measures $P,Q$ on a measurable space $(\cZ,\cA)$ is defined by $d(P,Q) \deq 2\sup_{A \in \cA} |P(A) - Q(A)|$. Denoting by $p$ and $q$ the respective densities of $P$ and $Q$ w.r.t. a dominating measure $\nu$, we can also write $d(P,Q) = \int_\cX |p(z) - q(z)| d\nu(z)$. The set of all $Q$ satisfying $d(P,Q) \le \delta$ for a given $P$ is called the {\em variational ball of radius $\delta$} around $P$.

Our first condition ensures that each source in the class is sufficiently close to an i.i.d. source, in an asymptotic sense. Define the $k$th $\beta$-mixing coefficient of $P_\theta$ \cite{Vid03} by\vspace{-4pt}
$$
\beta_\theta(k) \deq 2 \sup_{A \in \sigma(X^0_{-\infty},X^\infty_k)} |P_\theta(A) - P^-_\theta \times P^+_\theta (A) |,\vspace{-4pt}
$$
where $\sigma(X^0_{-\infty},X^\infty_k)$ is the $\sigma$-field generated by $\{X_i\}_{i \le 0}$ and $\{X_i\}_{i \ge k}$, and $P^-_\theta$ and $P^+_\theta$ are the marginal distributions of $\{X_i\}_{i \le 0}$ and $\{X_i\}_{i > 0}$, respectively. An i.i.d. source has $\beta(k) \equiv 0, \forall k$; if $\beta(k)\stackrel{k\to\infty}{\longrightarrow} 0$, the source is called {\em $\beta$-mixing}.

\noindent {\em Condition 1.} The sources in $\{P_\theta : \theta \in \Lambda\}$ are {\em algebraically $\beta$-mixing}:\vspace{-4pt}
$$
\exists r > 0 \mbox{ such that } \beta_\theta(k) = O(k^{-r}), \forall \theta \in \Lambda.\vspace{-4pt}
$$

The second condition ensures that the parametrization of the sources is sufficiently smooth.

\noindent {\em Condition 2.} Let $d_n(\theta,\theta')$ denote the variational distance between $P^n_\theta$ and $P^n_{\theta'}$. Then for every $\theta \in \Lambda$,\vspace{-4pt}
$$
\exists \delta_\theta,c_\theta > 0 \mbox{ such that } \sup_n \frac{d_n(\theta,\theta')}{\sqrt{n}} \le c_\theta \| \theta - \theta' \|\vspace{-4pt}
$$
for all $\theta'$ satisfying $\| \theta' - \theta \| < \delta_\theta$, where $\| \cdot \|$ denotes the Euclidean norm on $\R^k$.

This condition is met, for instance, if the asymptotic Fisher information matrix $I(\theta)$ exists for all $\theta \in \Lambda$ (under some technical assumptions on the densities $p^n_\theta$). It guarantees that, for every sequence $\{\delta_n\}_{n \in \N}$ of positive reals satisfying $\delta_n \to 0, \sqrt{n}\delta_n \to 0$ as $n \to \infty$, and for every sequence $\{\theta_n\}_{n \in \N}$ in $\Lambda$ satisfying $\| \theta_n - \theta \| < \delta_n$ for a given $\theta \in \Lambda$, we have $d_n(\theta_n,\theta) \to 0$ as $n \to \infty$. 

Finally, we impose a learnability condition. To state it we need some facts on Vapnik--Chervonenkis classes (see, e.g., \cite{DevLug01,Vid03}).  Let $(\cZ,\cA)$ be a measurable space. Given a collection $\cC$ of measurable subsets of $\cZ$, its {\em Vapnik-Chervonenkis (VC) dimension} $\sV(\cC)$ is defined as the largest integer $n$ for which\vspace{-4pt}
\begin{equation}
\max_{x^n \in \cX^n} |\{(1_{\{x_1 \in A\}},\cdots,1_{\{x_n \in A\}}) : A \in \cC \} | = 2^n;
\label{eq:vc}\vspace{-4pt}
\end{equation}
if (\ref{eq:vc}) holds for all $n$, then $\sV(\cC) = \infty$. If $\sV(\cC) < \infty$, we say that $\cC$ is a VC class. The Vapnik--Chervonenkis inequalities are finite-sample bounds on uniform deviations of probabilities of events in a VC class from their relative frequencies: if $X^n = (X_1,\cdots,X_n)$ is an i.i.d. sample from a probability measure $P$ on $(\cZ,\cA)$, and if $\cC$ is a VC class with $\sV(\cC) \ge 2$, then\vspace{-4pt}
$$
\Pr \Big\{\sup_{A \in \cC} |P_{X^n}(A) - P(A)| > \epsilon\Big\} \le 8n^{\sV(\cC)}e^{-n\epsilon^2/32}, \forall \epsilon > 0\vspace{-4pt}
$$
$$
\mbox{and }\E\Big\{\sup_{A \in \cC} |P_{X^n}(A) - P(A)| \Big\} \le c\sqrt{\sV(\cC) \log n/n},\vspace{-4pt}
$$
where $c > 0$ is a universal constant\footnote{Using more refined techniques, the $c\sqrt{\sV(\cC)\log n/n}$ bound can be improved to $c'\sqrt{\sV(\cC)/n}$, where $c'$ is another constant. However, $c'$ is much larger than $c$, so any benefit of the new bound shows only for ``impractically" large values of $n$.}, $P_{X^n}$ is the empirical distribution of $X^n$, and the probabilities and expectations are w.r.t. the product measure $P^n$ on $(\cZ^n,\cA^n)$.

\noindent {\em Condition 3.} For $n\in\N$, let $\cA_n$ consist of all sets of the form\vspace{-4pt}
$$
A_{\theta,\theta'} = \{x^n \in \cX^n : p_\theta(x^n) > p_{\theta'}(x^n) \}, \,\, \theta \neq \theta'\vspace{-4pt}
$$
($\cA_n$ is the so-called {\em Yatracos class} defined by $\{p^n_\theta\}$, see \cite{DevLug01} and references therein). Then we require that each $\cA_n$ is a VC class, $V_n \equiv \sV(\cA_n) < \infty$, and that $V_n = o(n/\log n)$.
\begin{theorem} \label{thm:main}
Suppose Conditions 1--3 are satisfied. Then for every $\lambda, \eta > 0$ there exists a sequence $\{C^{n,m_n}_*\}_{n \in \N}$ of variable-rate vector quantizers with memory lengths $m_n = n(n + \lceil n^{(2+\eta)/r}\rceil)$, such that\vspace{-4pt}
$$
L_\theta(C^{n,m_n}_*,\lambda) - \inf_m \inf_{C^{n,m}}
L_\theta(C^{n,m},\lambda) = O\left(\sqrt{\frac{V_n\log n}{n}}\right)\label{eq:universality}\vspace{-4pt}
$$
for all $\theta \in \Lambda$. Moreover, for each $n$, the binary description produced by the encoder is such that the decoder can identify the $n$-dimensional marginal of the active source up to a variational ball of radius $O\big(\sqrt{V_n \log n/n}\big)$ almost surely.\vspace{5pt}
\end{theorem}
That is, for each $n,\theta$ the code $C^{n,m_n}_*$, which is {\em independent of $\theta$}, performs almost as well as the best finite-memory quantizer with block length $n$ that can be designed with full knowledge of $P^n_\theta$. Thus, as far as compression goes, our scheme can compete with all finite-memory variable-rate quantizers, with the additional bonus of allowing the decoder to identify the active source in an asymptotically optimal manner. Recalling the discussion of Lagrangian optimality in Section~\ref{sec:prelims}, we see that Theorem~\ref{thm:main} immediately implies the following:
\begin{corollary} The sequence $\{C^{n,m_n}_*\}_{n \in \N}$ is {\em weakly minimax universal}\footnote{See \cite{ChoEffGra96} for other notions of universality for lossy codes.} for $\{P_\theta : \theta \in \Lambda\}$, i.e., for every $\theta \in \Lambda$, $L_{\theta_0}(C^{n,m_n}_*,\lambda) \to L_\theta(\lambda)$ as $n \to \infty$.
\end{corollary}

\section{The proof of Theorem~\ref{thm:main}}

\noindent {\em The main idea.} It suffices to construct a universal scheme that can compete with all {\em zero-memory} codes; that is, we need to show that there exists a sequence $\{C^{n,m_n}_*\}$ of codes, such that $L_\theta(C^{n,m_n}_*,\lambda) - \wh{L}^n_\theta(\lambda) = O(\sqrt{V_n \log n/n})$ for all $\theta \in \Lambda$.

We assume throughout that the ``true" source is $P_{\theta_0}$ for some $\theta_0 \in \Lambda$. Our code operates as follows. Suppose that both the encoder and the decoder have access to a countably infinite ``database" $\bd{c} = \{\theta(i)\}_{i \in \N} \subset \Lambda$. Using Elias' universal representation of the integers \cite{Eli75}, we can associate to each $\theta(i)$ a unique binary string $s(i)$ with $\ell(s(i)) = \log i + O(\log \log i)$ bits.
 Suppose also that for each $n,\theta$ there exists a zero-memory $n$-block code $C^n_\theta = (f_\theta,\varphi_\theta)$ that achieves the $n$th-order Lagrangian optimum for $P_\theta$: $L_\theta(C^n_\theta,\lambda) = \wh{L}^n_\theta(\lambda)$. The encoding of $X^n_1$ into $\hX^n_1$ is done as follows:
\begin{enumerate}
\item The encoder estimates $P^n_{\theta_0}$ from the $m_n$-block $X^0_{-m_n+1}$ as $P^n_{\td{\theta}}$, where $\td{\theta} = \td{\theta}(X^0_{-m_n+1})$.
\item The encoder then computes the {\em waiting time}\vspace{-4pt}
$$
T_n \deq \inf \big\{ i \ge 1 : d_n\big(\theta(i),\td{\theta}(X^0_{-m_n+1})\big) \le \sqrt{n}\delta_n \big\},\vspace{-4pt}
$$
with the standard convention that the infimum of the empty set is equal to $+\infty$; $\{\delta_n\}$ is a sequence of positive reals to be specified later.
\item If $T_n < + \infty$, the encoder sets $\wh{\theta} = \theta(T_n)$; otherwise, the encoder sets $\wh{\theta} = \theta(1)$ (or some other default $\theta$).
\item The description of $X^n_1$ is a concatenation of three binary strings: (i) a 1-bit flag $b$ to tell whether $T_n$ is finite $(b=0)$ or infinite $(b=1)$; (ii) a binary string $s_1$ which is equal to $s(T_n)$ if $T_n < +\infty$ or is empty if $T_n = +\infty$; (iii) $s_2 = f_{\wh{\theta}}(X^n_1)$. The string $\td{s} = bs_1$ is the {\em first-stage description}, while $s_2$ is the {\em second-stage description}.
\end{enumerate}
The decoder receives $bs_1s_2$, determines $\wh{\theta}$ from $\td{s}$, and produces $\wh{X}^n_1 = \varphi_{\wh{\theta}}(s)$. If $b = 0$ (which, as we shall show, will happen eventually a.s.), then $P^n_{\wh{\theta}}$ is in the variational ball of radius $\sqrt{n}\delta_n$ around the estimated $P^n_{\td{\theta}}$. If the latter is a good estimate, i.e., $d_n(\theta_0,\td{\theta}) \stackrel{{\rm a.s.}}{\to} 0$ as $n \to \infty$, then the decoder's estimate of $P^n_{\theta_0}$ is only slightly worse. Moreover, the a.s. convergence of $d_n(\theta_0,\wh{\theta})$ to zero as $n \to \infty$ implies that the performance of $C^n_{\wh{\theta}}$ on $P_{\theta_0}$ is close to the optimum $L_{\theta_0}(C^n_{\theta_0},\lambda) \equiv \wh{L}^n_{\theta_0}(\lambda)$.

Formally, the code $C^{n,m_n}_*$ is comprised by the following maps: (1) the {\em parameter estimator} $\map{\td{\theta}}{\cX^{m_n}}{\Lambda}$; (2) the {\em parameter encoder} $\map{\td{g}}{\Lambda}{\td{\cS}}$, where $\td{\cS} = \{ 0 s(i) \}_{i \in \N} \cup \{1 \}$; (3) the {\em parameter decoder} $\map{\td{\psi}}{\td{\cS}}{\Lambda}$. Let $\td{f}$ denote the composition $\td{g}\circ\td{\theta}$ of the parameter estimator and the parameter encoder, which we refer to as the {\em first-stage encoder}, and let $\wh{\theta}$ denote the composition $\td{\psi} \circ \td{f}$ of the parameter decoder and the first-stage encoder. The decoder $\td{\psi}$ is the {\em first-stage decoder}. The collection $\{C^n_\theta : \theta \in \Lambda\}$ defines the {\em second-stage codes}. The encoder $\map{f_*}{\cX^n \times \cX^{m_n}}{\td{\cS} \times \cS}$ and the decoder $\map{\varphi_*}{\td{\cS} \times \cS}{\hcX^n}$ of $C^{n,m_n}_*$ are defined as $f_*(X^n_1,X^0_{-m_n+1}) = \td{f}(X^0_{-m_n+1})f_{\wh{\theta}(X^0_{-m_n+1})}(X^n_1)$ and $\varphi_*(\td{s}s) = \varphi_{\td{\psi}(\td{s})}(s)$ for all $s \in \cS, \td{s} \in \td{\cS}$, respectively.

\begin{figure}
\centerline{
\includegraphics[width=0.9\columnwidth]{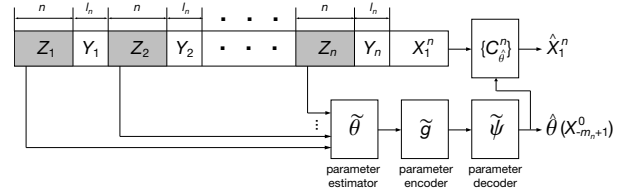}
}\vspace{-4pt}
\caption{The structure of the code $C^{n,m_n}_*$. The shaded blocks are those used for estimating the source parameters.}\label{fig:code_structure}\vspace{-15pt}
\end{figure}

To assess the performance of the code, introduce the functions $g(x^n,y^{m_n}) = \rho_n(x^n,C^n_{\wh{\theta}(y^{m_n})}(x^n)) + \lambda \ell_n ( f_{\wh{\theta}(y^{m_n})}(x^n))$ and $h(y^{m_n}) = \ell_n ( \td{f}(y^{m_n}) )$. Then $h(X^0_{-m_n+1})$ is the normalized length of the first-stage description, while $g(X^n_1,X^0_{-m_n+1})$ is the instantaneous Lagrangian performance of the corresponding second-stage code. The expected Lagrangian performance of our code is\vspace{-4pt}
$$
L_{\theta_0}(C^{n,m_n}_*,\lambda) = \E_{\theta_0} g(X^n_1,X^0_{-m_n+1}) + \lambda \E_{\theta_0} h(X^0_{-m_n+1}).\vspace{-4pt}
$$
We prove the theorem by showing that, with proper choices for the memory length $m_n$, the ``database" $\bd{c}$, the parameter estimator $\td{\theta}$, and the sequence $\{\delta_n\}$, we can ensure that $\E_{\theta_0} h(X^0_{-m_n+1}) = O(k\log n/n) + O(\log\log n/n) + o(1)$, $\E_{\theta_0} g(X^n_1,X^0_{-m_n+1}) = \wh{L}^n_\theta(\lambda) + O(\sqrt{V_n\log n/n})$, and $d_n(\theta_0,\wh{\theta}(X^0_{-m_n+1})) = O(\sqrt{V_n\log n/n})$ $P_{\theta_0}$-almost surely.

\noindent {\em Step 1: choice of memory length.} Let $l_n = \lceil n^{(2+\eta)/r} \rceil$ and $m_n = n(n+l_n)$. Divide $X^0_{-m_n+1}$ into $n$ blocks $Z_1,\ldots,Z_n$ of length $n$ interleaved by $n$ blocks $Y_1,\ldots,Y_n$ of length $l_n$ (see Figure~\ref{fig:code_structure}). The parameter estimator $\td{\theta}$, although defined as acting on the entire $X^0_{-m_n+1}$, effectively will make use only of $Z^n = (Z_1,\ldots,Z_n)$. Each $Z_j \sim P^n_{\theta_0}$, but the $Z_j$'s are not independent. Let $Q^{(n)}$ denote the marginal distribution of $Z^n$, and let $\td{Q}^{(n)}$ denote the product of $n$ copies of $P^n_{\theta_0}$. Using induction and the definition of the $\beta$-mixing coefficient, we can show that $\dvar (Q^{(n)},\td{Q}^{(n)}) \le (n-1) \beta_{\theta_0}(l_n) = O(1/n^{1+\eta})$, which follows from Condition 1 and our choice of $l_n$. This ``blocking technique" \cite{Yu94} allows us to approximate certain probabilities and expectations w.r.t.  $P_{\theta_0}$ by probabilities and expectations w.r.t. suitably constructed i.i.d. processes.

\noindent {\em Step 2: construction of the database.} We proceed by random selection. Let $W$ be some probability measure on $\Lambda$ with a positive, everywhere continuous density $w(\theta)$. We generate $\bd{C} = \{\theta(i)\}_{i \in \N}$ as an i.i.d. sequence of vectors in $\Lambda$ drawn according to $W$, independently of $\bd{X}$.

\noindent {\em Step 3: estimation of the active source.} We use the Devroye--Lugosi {\em minimum-distance estimator} (MDE) (see \cite{DevLug01} and references therein). Namely, given the estimation blocks $Z^n = (Z_1,\ldots,Z_n)$, define $U_\theta(Z^n) \deq \sup_{A \in \cA_n} |P^n_\theta(A) - P_{Z^n}(A)|$ for every $\theta \in \Lambda$, where the supremum is over all sets in the Yatracos class $\cA_n$ and $P_{Z^n}$ is the empirical distribution on $\cX^n$ induced by $Z^n$. Then $\td{\theta}(X^0_{-m_n+1})$ is any $\theta^* \in \Lambda$ satisfying $U_{\theta^*}(Z^n_1) < \inf_{\theta \in \Lambda} U_\theta(Z^n_1) + 1/n$ (the extra $1/n$ term ensures that at least one such $\theta^*$ exists). Note that $\td{\theta}(X^0_{-m_n+1})$ only depends on $Z^n$. The key property of the MDE is  \cite{DevLug01}\vspace{-3pt}
\begin{equation}
d_n(\theta_0,\td{\theta}(X^0_{-m_n+1})) \le 4U_{\theta_0}(Z^n_1) + 3/n,
\label{eq:mde_property}\vspace{-3pt}
\end{equation}
which holds regardless of whether $Z^n$ is i.i.d. or not.

\noindent {\em Step 4: expected first-stage description length.} We follow the ideas of \cite{KonZha02}. Let us assume that the sequence $\{\delta_n\}$ is such that $\delta_n \to 0$ as $n\to\infty$. Define the event $F_n = \{ \theta \in \Lambda: d_n(\theta,\td{\theta}(X^0_{-m_n+1})) \le \sqrt{n}\delta_n \}$ and note that if $q_n = W(F_n|X^0_{-m_n+1} = x^0_{-m_n+1}) > 0$, then the waiting time $T_n$ is a geometric random variable with parameter $q_n$. Condition~2 ensures that, in fact, $q_n > 0$ for $n$ sufficiently large, for $P_{\theta_0}$-almost all realizations of $\bd{X}$. Using the Borel--Cantelli lemma, it is not hard to show that $\E_{\theta_0}  \log T_n  \le \log \log n + 2 - \E_{\theta_0} \log q_n $ for all realizations of $\bd{C}$, eventually $P_{\theta_0}$-a.s. We now lower-bound $q_n$ for large $n$. Using the triangle inequality, independence of $\bd{X}$ and $\bd{C}$, Condition 2 and the fact that $\delta_n \to 0$ as $n \to \infty$, we have, for $n$ sufficiently large,\vspace{-4pt}
$$
q_n \ge W\Big( \| \Theta - \theta_0 \| \le \delta_n/2c_{\theta_0}\Big) P_{\theta_0}\Big(d_n(\theta_0,\td{\theta}) \le \sqrt{n}\delta_n/2 \Big),\vspace{-4pt}
$$
where $\td{\theta} = \td{\theta}(X^0_{-m_n+1})$ and $\Theta \sim W$. Via simple volume bounding, $ W\big( \| \Theta - \theta_0 \| \le \delta_n/2c_{\theta_0}\big) \ge (1/2) w(\theta_0) v_k (\delta_n/2c_{\theta_0})^k$ for $n$ sufficiently large, where $v_k$ is the volume of the unit sphere in $\R^k$. Next, we use blocking to approximate $P_{\theta_0}$-probabilities by $\td{Q}^{(n)}$-probabilities, and then invoke the property (\ref{eq:mde_property}) of the MDE and the Vapnik--Chervonenkis inequalities to obtain\vspace{-4pt}
\begin{eqnarray*}
&& P_{\theta_0}\Big(d_n(\theta_0,\td{\theta}(X^0_{-m_n+1})) \le \sqrt{n}\delta_n/2 \Big) \nonumber \\
&& \quad \ge 1 - 8n^{\sV(\cA_n)}e^{-n(\sqrt{n}\delta_n - 6/n)^2/2048} - O(1/n^{1+\eta}).\vspace{-4pt}
\end{eqnarray*}
Choosing $\delta_n = \frac{\sqrt{2048 (V_n+1)\ln n}}{n} + \frac{6}{n^{3/2}}$, we get for the normalized expected first-stage description length\footnote{Note that, up to a constant, the first term on the right-hand side has the same form as in Rissanen \cite{Ris84}; additional terms are due to the unboundedness of $\Lambda$ and the fact that the points $\theta(i)$ do not form a regular grid.}\vspace{-4pt}
$$
\E_{\theta_0} h(X^0_{-m_n+1}) = O(k\log n/n) + O(\log \log n/n) + o(1).\vspace{-4pt}
$$
The sequence $\delta_n$ indeed converges to 0 owing to Condition~3.
\noindent {\em Step 5: expected second-stage Lagrangian performance.} Using the fact that the distortion measure $\rho$ is bounded, one can show via an argument similar to the proof of Lemma~9 in Section~10 of \cite{Lin01} that for every $\theta \in \Lambda$ there is no loss of generality in assuming that an $n$-block code $C^n_\theta = (f_\theta,\varphi_\theta)$ achieving $\wh{L}^n_\theta(\lambda)$ satisfies $\ell_n(f_\theta(x^n)) \le 2\rho_{\max}/\lambda$ for all $x^n \in \cX^n$. Thus, $g$ is bounded by $3\rho_{\max}$. A straightforward application of Fubini's theorem and the definition of the $\beta$-mixing coefficient yields $\E_{\theta_0} g(X^n_1,X^0_{-m_n+1}) \le \E_{\theta_0} L_{\theta_0}(C^n_{\wh{\theta}},\lambda) + O(1/n^{2+\eta})$, where $\wh{\theta} = \wh{\theta}(X^0_{-m_n+1})$. Thus, the Lagrangian performance of the second-stage code is determined by the behavior of the code $C^n_{\wh{\theta}}$ (which depends on $X^0_{-m_n+1}$). Because $\rho$ is a metric, a basic Lagrangian mismatch argument (see, e.g., Lemma~9 in Section~8 of \cite{Lin01}) shows that\vspace{-4pt}
$$
\E_{\theta_0} L_{\theta_0}(C^n_{\wh{\theta}},\lambda) \le \E_{\theta_0} L_{\theta_0}(C^n_{\theta_0},\lambda) + 4\rho_{\max}\E_{\theta_0} d_n(\theta_0,\wh{\theta}).\vspace{-4pt}
$$
By blocking, the expectation of $d_n(\theta_0,\wh{\theta})$ w.r.t. $P_{\theta_0}$ can be approximated by expectation w.r.t. $\td{Q}^{(n)}$. Followed by an application of the triangle inequality, this yields\vspace{-3pt}
$$
\E_{\theta_0} d_n(\theta_0,\wh{\theta}) \le \E_{\td{Q}^{(n)}} \big\{ d_n(\theta_0,\td{\theta}) + d_n(\td{\theta},\wh{\theta}) \big \} + O(1/n^{1+\eta}),\vspace{-3pt}
$$
where $\td{\theta} = \td{\theta}(X^0_{-m_n+1})$ is the MD estimate of $\theta_0$. Now, $d_n(\td{\theta},\wh{\theta}) \le \sqrt{n}\delta_n = O(\sqrt{V_n \log n/n})$ eventually almost surely, by construction of the first-stage encoder. The expectation $\E_{\td{Q}^{(n)}} d_n(\theta_0,\td{\theta})$ can be handled via (\ref{eq:mde_property}) and the Vapnik--Chervonenkis inequalities, yielding\vspace{-2pt}
$$
\E_{\theta_0} d_n(\theta_0,\td{\theta}) = O\big(\sqrt{V_n \log n/n}\big) + O(1/n^{1+\eta}).\vspace{-4pt}
$$
$$
\mbox{Thus, }\E_{\theta_0} \{g\} = \wh{L}^n_{\theta_0}(\lambda) + O(\sqrt{V_n \log n/n}) + O(1/n^{1+\eta}).\vspace{-4pt}
$$
\noindent {\em Step 6: the overall performance.} Gathering together our estimates for the first stage and for the second stage, we get
\begin{eqnarray*}
&& L_{\theta_0}(C^{n,m_n}_*,\lambda) = \wh{L}^n_{\theta_0}(\lambda) + O(\sqrt{V_n \log n/n}) \\
&& \quad \quad + O(k\log n/n) + O(\log\log n/n) + o(1)
\end{eqnarray*}
for almost every realization of the database $\bd{C}$. As for the performance of the scheme in identifying the active source, note that, with our choice of $l_n$, the sequence $n\beta_{\theta_0}(l_n)$ is summable in $n$. Then a straightforward application of the Borel--Cantelli lemma and the Vapnik--Chervonenkis inequalities yields\vspace{-4pt}
$$
d_n\left(\theta_0,\wh{\theta}(X^0_{-m_n+1})\right) = O\left(\sqrt{V_n \log n/n}\right), \,\, P_{\theta_0}-\mbox{a.s.}.
$$

\section{Examples}
\label{sec:examples}

Here, we present three examples of parametric families
satisfying the conditions of Theorem~\ref{thm:main} and thus admitting
joint universal lossy coding and identification schemes. The following   result \cite{Vid03} will be used throughout: Let $\cC = \{A_\xi : \xi \in \R^N\}$ be a collection of measurable subsets of $\R^d$, such that $A_\xi = \{ z \in \R^d : \Pi(z,\xi) > 0 \}$ for all $\xi$, where for each $z \in \R^d$, $\Pi(z,\cdot)$ is a polynomial of degree $s$ in the components of $\xi$. Then $\cC$ is a VC class with $\sV(\cC) \le 2N\log(4es)$. 

\noindent {\bf Stationary memoryless sources.} Let $\cX = \R$, and let $\{P_\theta : \theta \in \Lambda\}$ be the collection of all Gaussian i.i.d. processes with mean $m \in \R$ and variance $\sigma \in (0,\infty)$. Thus $\Lambda = \{(m,\sigma) : m \in \R, 0 < \sigma < \infty \} \subset \R^2$. This class of sources trivially satisfies Condition 1 with $r = +\infty$, and it remains to check Conditions 2 and 3. To check Condition 2, consider the normalized relative entropy (information divergence) $D_n(\theta \| \theta')$ between $P^n_\theta$ and $P^n_{\theta'}$, with $\theta = (m,\sigma)$ and $\theta' = (m',\sigma')$ (which is equal to $D_1(\theta \| \theta')$ because the sources are i.i.d.). It is not hard to get the bound
$D_n(\theta \| {\theta'}) \le \left(1 + \sigma'/\sigma\right)^2  \| \theta - \theta' \|^2/2{\sigma'}^2$. Now fix a small $\delta \in (0,\sigma)$ and suppose that $\| \theta - \theta' \|  < \delta$. Then $|\sigma - \sigma'| < \delta$, so we can further upper-bound $D_n(\theta \| {\theta'})$ as $D_n(\theta \| {\theta'}) \le \frac{c^2_\theta}{2} \| \theta - \theta'\|^2$ for all $\theta'$ in the open ball of radius $\delta$ around $\theta$, with $c_\theta = 3/(\sigma - \delta)$. Using Pinsker's inequality \cite{DevLug01}, we have $d_n(\theta,\theta')/\sqrt{n} \le \sqrt{2 D_n(\theta \| {\theta'})} \le c_\theta \| \theta - \theta' ||$ for all $n$. Thus, Condition 2 holds. To check Condition 3 note that, for each $n$, the Yatracos class $\cA_n$ consists of all sets of the form
$\left\{ x^n \in \R^n : \Pi(x^n,\theta,\theta') > 0 \right\}$, $\theta,\theta' \in \Lambda$, where for each $x^n \in \cX^n$ $\Pi(x^n,\theta,\theta')$ is a third-degree polynomial in $(\ln \sigma^2, \ln {\sigma'}^2,1/\sigma^2,1/{\sigma'}^2,m,m')$. Thus, $\cA_n$ is a VC class with $\sV(\cA_n) \le 12\log(12e)$, satisfying Condition~3.

\noindent {\bf Autoregressive (AR) sources.} Let $\cX = \R$ and let $\bd{X}$ be a Gaussian AR($p$) source. That is, there exist $p$ real parameters $a_1,\ldots,a_p$, such that $X_n = - \sum^p_{i=1} a_i X_{n-i} + Y_n$ for all $n$, where $\bd{Y} = \{Y_i\}_{i \in \Z}$ is an i.i.d. Gaussian process with zero mean and unit variance. Let $\Lambda \subset \R^p$ be the set of all $a_1,\ldots,a_p$, such that all roots of the  polynomial $A(z) = \sum^p_{i=0}a_i z^i$, $a_0 \equiv 1$, lie outside the unit circle in the complex plane. Under these conditions, for each $\theta \in \Lambda$ the process $\bd{X}$ is {\em exponentially $\beta$-mixing} \cite{Mok88}, i.e., there exists some $\gamma = \gamma(\theta) \in (0,1)$, such that $\beta_\theta(k) = O(\gamma^k)$. Now, for any fixed $r > 0$, $\gamma^k \le k^{-r}$ for $k$ sufficiently large, so Condition 1 holds. For Condition 2, it can be shown that, for each $\theta \in \Lambda$, the asymptotic Fisher information matrix $I(\theta)$ exists (and is nonsingular) \cite{KleSpr06}. Thus, Condition 2 can be met. To verify Condition 3, consider the $n$-dimensional marginal $P_\theta(x^n)$, which has the normal density
$p_\theta(x^n) = \cN(x^n; 0, R_n(\theta))$, where $R_n(\theta)$ is the $n$th-order autocorrelation matrix of $\bd{X}$. For every $\theta \in \Lambda$, let $\bar{\theta} = (\theta,\ln \det R^{-1}_n(\theta))$. Since $\ln \det R^{-1}_n(\theta)$ is uniquely determined by $\theta$, we have $A_{\theta,\theta'} = A_{\bar{\theta},\bar{\theta}'}$ for all sets in the Yatracos class $\cA_n$. This, and the fact that the entries of $R^{-1}_n(\theta)$ are quadratic functions of $a_1,\ldots,a_p$, implies that, for each $x^n$, the condition $x^n \in A_{\theta,\theta'}$ can be expressed as $\Pi(x^n,\bar{\theta},\bar{\theta}') > 0$, where $\Pi(x^n,\cdot)$ is quadratic in the $2p+2$ real variables $\bar{\theta}_1,\ldots,\bar{\theta}_{p+1},\bar{\theta}'_1,\ldots,\bar{\theta}'_{p+1}$. Thus, $\sV(\cA_n) \le (4p+4)\log(8e)$. Therefore, Condition 3 is met.

\noindent {\bf Hidden Markov processes.} A hidden Markov process is a discrete-time finite-state homogeneous Markov chain, observed through a discrete-time memoryless channel (see \cite{EphMer02} and references therein). Let $\bd{S} = \{S_i\}_{i \in \Z}$ be a stationary ergodic Markov process with $M < \infty$ states and the (unique) stationary distribution $\pi = (\pi_1,\ldots,\pi_M)$. Let $a_{ij} = \Pr(S_{t+1} = j | S_t = i)$, $1 \le i,j \le M$, denote the corresponding one-step transition probabilities. Let $\cX = \R^d$, and consider a discrete-time memoryless channel with input alphabet $\cS \deq \{1,\ldots,M\}$ and output alphabet $\cX$, specified by a collection $\{ p(\cdot|s) : s \in \cS \}$ of probability densities on $\R^d$ w.r.t. the Lebesgue measure. The output process $\bd{X} = \{X_i\}_{i \in \Z}$ is the source of interest.

Let us assume that the channel transition densities are known, and that the one-step transition probabilities of the underlying Markov chain $\bd{S}$ are known to be strictly positive and bounded from below by some $a_0 > 0$. Thus, our parameter space is the set $\Lambda = \left\{ \theta = [a_{ij}] \in \R^{M \times M} : a_{ij} > a_0, \forall i,j \right\}$. Under these assumptions, for any $\theta \in \Lambda$ the underlying Markov process $\bd{S}$ is exponentially $\beta$-mixing \cite{Bil95}. It can also be shown \cite{Vid03} that for every $\theta \in \Lambda$ there exists a measurable map $\map{F}{\cS \times [0,1]}{\cX}$, such that $X_i = F(S_i,U_i)$ for all $i \in \Z$, where $U_i$ are i.i.d. random variables with uniform distribution on $[0,1]$, independent of $\bd{S}$. The pair process $\{(S_i,U_i)\}$ is exponentially $\beta$-mixing, and therefore so is $\bd{X}$. This establishes Condition~1. Under additional technical assumptions on the densities $\{p(\cdot|s)\}$ it can be shown that the asymptotic Fisher information matrix $I(\theta)$ exists for all $\theta \in \Lambda$ \cite{DouMouRyd04}, which implies that Condition~2 holds as well. Finally, to show that Condition~3 is satisfied, note that the $n$-dimensional marginal of $P_{\theta}$ for a given $\theta = [a_{ij}]$ has the density $p_\theta(x^n) = \sum_{s^n \in \cS^n} \prod^n_{i=1} a_{s_{i - 1}s_i}p(x_i|s_i)$, where $a_{s_0s} \equiv \pi_s$ for all $s$. Then it follows that the Yatracos class $\cA_n$ consists of sets of the form $\{x^n \in \cX^n : \Pi(x^n,\theta,\theta') > 0\}$, $\theta = [a_{ij}], \theta' = [a'_{ij}] \in \Lambda$, where $\Pi(x^n,\cdot)$ is a polynomial of degree $n$ in the $2M^2$ parameters $\{a_{ij},a'_{ij}\}$. Thus, $\sV(\cA_n) \le 4M^2\log(4en)$, so that Condition~3 holds as well.
\section*{Acknowledgment}

The author would like to thank Andrew Barron, Ioannis Kontoyiannis and Mokshay Madiman for useful discussions. This work was supported by the Beckman Fellowship. 

\bibliography{mixing_isit2007}

\end{document}